# *A Modified T-Mixer Simulation, Fabrication, and Characterization For 1D Diffusion Controlled Studies*

M. Rashed, Khan[1,*], Taliman Afroz[2], Caleb Shaw[1]

[1]Department of Chemical and Materials Engineering, University of Nevada, Reno; Nevada, U.S.A.

[2]Personalized Miniaturized Systems (PMS) LLC, Reno, Nevada, U. S. A.

[#]Corresponding Author: M. Rashed Khan mrkhan@unr.edu

**Abstract**

Here, we describe a simple and unique architecture of a microfluidic mixer that can mix two streams (Water and Fluorescein Isothiocyanate (FITC) buffer) with 30% efficiency. The overall mixing in this design is dominated by 1D diffusion, and to enhance mixing, we used three different types of geometric obstacles inside the channel. Comsol multiphysics simulation software was used to validate the theoretical mixing efficiency (at Reynolds Number, Re ~ 0.1, 1, and 10) of this device. We utilized soft lithography and replica molding techniques to fabricate the device out of Polydimethylsiloxane (PDMS, a commonly used polymer) on a glass substrate. The effective length of our microfluidic mixer is 5mm, and the channel width is about 200 microns with 50 micron height. It is composed of three different shapes of obstacles (e.g., 6 cones, 3 arrays of rectangular bars, and 5 circular posts), and all of these are placed inside the main channel. FITC buffer (Diffusion co-efficient, D = 0.5 x $10^{-9}$ $m^2$/s) and DI water were used to investigate the mixer performance at Re~1. Simulated and experimental results are based on approximately 1.1 $mm^2$ flow area and suggest that 30% mixing is achievable with the current design.

# Introduction

Microfluidics integrates several different operations, such as mixing, chemical or biochemical reactions, separation, detection, and analysis in a single device in which one of the dimensions is on the sub-millimeter scale [1-6]. As different analytical operations can be carried out in a single device, it is exponentially growing as an attractive approach to different research areas where fluid manipulation, control, and actuation are needed at the micro or even nanoscale range [3,5]. Apart from the numerous advantages that microfluidics offer, a few issues related to active fluid flow in microchannel or temperature measurements/controls arise during micromixing [7-8]. The effective surface forces become the dominating parameters to control and manipulate fluids [2]. In this small length scale, flow is generally characterized by the low Reynolds number (Re), whereas the 1D slow diffusion process is the key to mixing fluid streams [2-3].

In contrast, mixing at the macro scale can be attained by flowing fluids at a high Reynolds number or integrating external stimuli that cause turbulence and enhance mixing [1, 2, 7]. However, turbulent mixing is not possible at microscales where the fluid flow is mostly laminar [1-3, 7]. As a result, molecular diffusion drives mixing, which is generally a slow process [2]. It is often desirable to accelerate this slow diffusion process by increasing the temperature, which will increase the diffusivity, utilizing a higher gradient in concentrations or increasing the area of the contacting fluids. In most cases, micromixers are made from polymeric materials, and these materials are not good thermal conductors. Increasing temperature to enhance diffusivity is thus very hard to achieve. To achieve efficient mixing, micromixers generally utilize a higher contact area or larger gradient in concentration between incoming streams [1-6, 8-9].

Micromixers are generally categorized as active or passive based on the choice and utilization of external energy as a source to enhance mixing [1, 2, 10-12]. Active mixer integrates external forces so that the time-dependent perturbations overcome the sluggish nature of the 1D molecular diffusion process [2]. Higher mixing efficiency can be achieved via active mixing of fluid streams, but these mixers are hard to fabricate, prone to irreversible damage, and have the possibility of degrading biological fluids as well. Passive micromixers are, therefore, the popular choice in recent years that typically rely on the pressure-driven flow of fluids and geometric construction of the microdevice [1, 2, 7- 13].

Passive micromixers rely heavily on the architectures of the microdevice to reduce the diffusion length [2, 7, 10, 13]. As diffusion length varies with square power, a small decrease in the mixing length can reduce the time for mixing as well as increase the area of contacting fluid. Most of the passive micromixers utilize different geometry to enhance the advection and the mass transport between streams [2]. Depending on the use of passive mixers, numerous designs are possible, and to date, a wide spectrum of designs has already been reported in the literature [1-29]. Some of the typical approaches to enhance the mixing efficiency of passive mixers include lamination, hydrodynamic focusing, mixing inside droplets, utilizing chaotic advection, intermittent solute injection, and porous structures in a packed bed [1, 8- 10, 13-29]. All these approaches are typically more advanced than a simple T mixer or Y mixer because of the nature of mixing requirements, and typically, the mixing efficiency is relatively lower for a T/Y mixer. In contrast, the device architecture of these advanced passive mixers is complex and very hard to implement [8-10]. A T mixer with different geometric constrictions is simple, easy to fabricate, and faster to carry out measurements and analysis. In-plane simple passive mixers avoid the issues of micromachining to fabricate complex 3D structures. Tesla or modified Tesla mixers [15] are the best examples of in-plane simple passive mixers, but here, our idea is to design a simple mixer that can recombine and split fluid in an area of 1.1 mm$^2$. We have used a novel architecture where we have integrated different shaped obstacles so that the incoming streams are separated into several sub-streams. We hypothesize that a simple micromixer with the option of splitting streams can achieve larger mixing efficiency when we put two or three of the same units in a series. We have only considered a 5mm length for our mixer, but in reality, it can go up to a centimeter in scale for better efficiency.

**Concept and Simulation**

We took a simple T-mixer [16] as our reference design and then incorporated geometric micro-obstacles into this T-mixer to split up the main streams of two fluids into several sub-streams. The idea is to fold the flow path, stretch out the diffusive width, and mix fluids down the length of the channel in a simple fashion [2]. The final architecture of the current micromixer is an optimum microfluidic model based on two different physics (e.g., laminar flow and 1D transport diffusion of dilute species) that are linked together. For a 200μm x 50μm x 7mm channel, initially, we picked 3 different factors (e.g., the ratio of channel opening, array of bars in the channels, and number of circular obstacles in between bars) and varied each of the factors by 5 sets of variables. For a fixed Reynolds number (i.e., 1 initially), we have, therefore, 15 different designs to run to find our optimum microfluidic mixer. Table 1 shows all the factors and the variables we have used to run the simulation.

Table 1: All the factors and variables for the current micromixer that are tested in Comsol for optimum design

| Factors | | | | | |
|---|---|---|---|---|---|
| Ratio | 50:50 | 100:50 | 150:50 | 200:50 | 250:50 |
| Array of Bars | 0 | 1 | 2 | 3 | 4 |
| Circular obstacle | 1 | 2 | 3 | 4 | 5 |

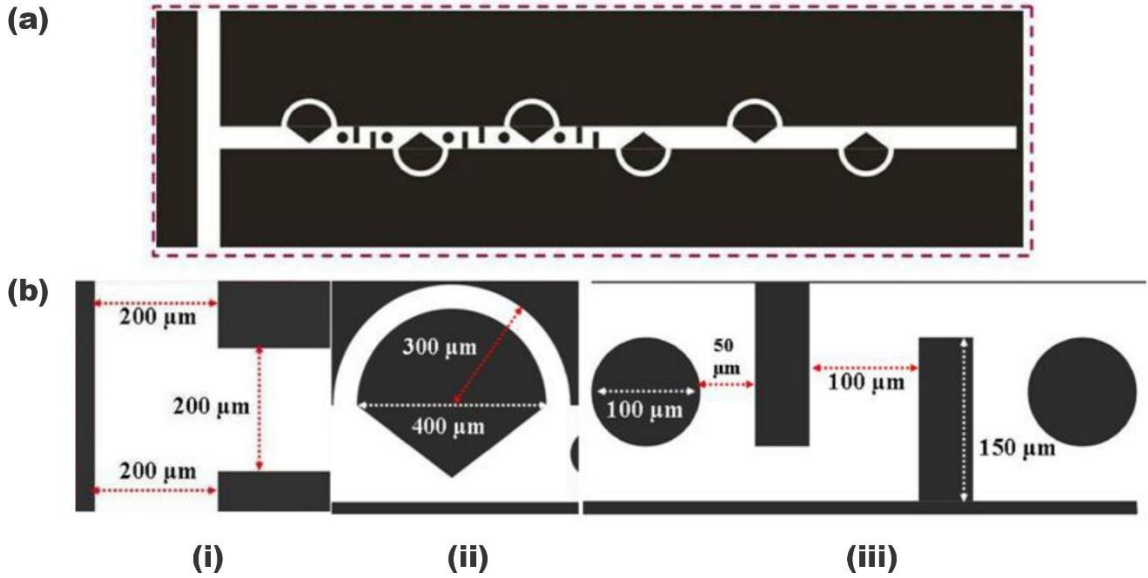

**F**igure 1: Idea of the design. (a) The schematic of the design shows all the obstacles inside the channel. (b) Dimensions of the obstacles. The frequency of placing the cone inside the channel is 1000µm. (i) Dimension of the inlet. (ii) Typical dimension of the cone. (iii) Dimensions of the rectangular bars and circular obstacles.

Figure 1a shows the schematic of our design. The dimensions of the two inlet channels are 200µm x 50µm x 7mm, which is shown in (i) of Figure 1(b). We made the length of the inlets sufficiently long to avoid any initial time perturbation in mixing. Then, we introduced 5 cone-shaped obstacles in the mixing region of our design as indicated by (ii) of Figure 1b, and we added one extra cone-shaped obstacle after 5mm, which acts like a buffer to remove experimental complexity. The distance between two successive cones is 1000µm (center to center), and the sharp edges of the cones meet at an angle of 60°. We introduce our first factor as the ratio of channel opening in between the circumferential width at the cone and the gap between the sharp edges of the wall. The half-circular path acts like a recirculation line to split streams, and it will recombine once the fluid re-enters the main channel. We introduce an array of bars as shown in (iii) of Figure 1(b), which is composed of two bars of 150µm x 50µm, and the distance between two bars is 100µm. The number of arrays varies from 0 to 4, and each array is placed between two cones. The last set of designs, where we introduce a 100µm circular obstacle in between bars and cones, is shown in (iii) of Figure 1(b). The gap between the bar and circles is 50µm. The idea behind splitting main streams of fluid into several subsequent sub-streams is to accelerate t diffusive mixing.

To test our device theoretically, we used COMSOL multiphysics finite element simulation software. Laminar fluid flow and transport diffusion physics are inked in this program. We set the inlet velocity at 0.013 m/s, which corresponds to Re ~1. After completing the device drawing with the first level of factors, we meshed the whole flow area into 80,000 to 100,000 free triangular areas. For meshing, we have used free triangular patterns with two overall refinements. The walls, corners, and sharp edges are meshed finely using selective meshing as well. We ran our simulation until the error reached the $10^{-6}$ range.

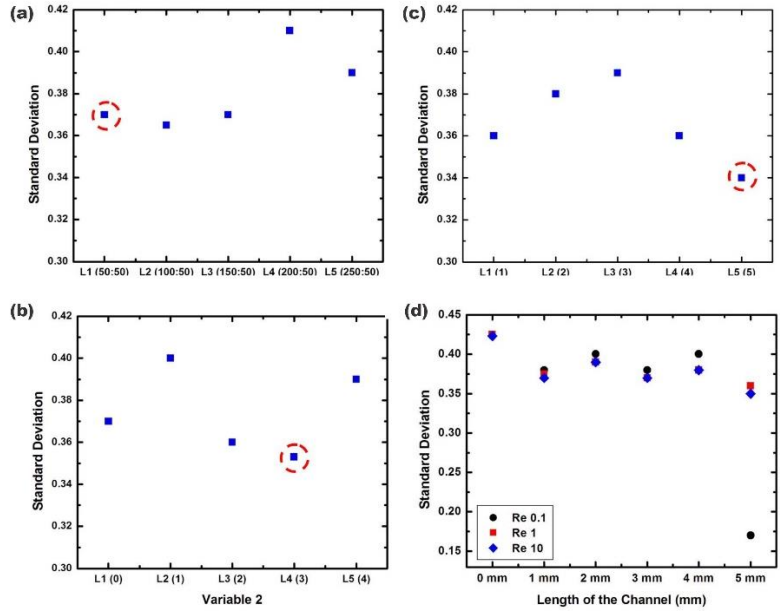

**Figure 2**: Optimization of the design by simulation in COMSOL multiphysics. Red circles indicate all the optimized numbers. (a) Optimization of variable 1, Ratio, (b) Optimization of variable 2, Array of bars, (c) Optimization of variable 3, Number of circular obstacles, (d) Simulation at different Reynolds numbers after optimizing variables 1, 2, 3.

All the simulation plots are shown in Figure 2. We focus on estimating the standard deviation as a measure of finding mixing efficiency (e.g., Mixing efficiency, $E = (1 - 2\sigma) \times 100$). A deviation of 0.5 indicates no mixing in the channel, while a value of 0 indicates complete mixing. We changed the ratio from 1:1 (i.e., 50μm wide recirculation around the cone and 50μm opening between the wall and sharp edge) to 5:1. The opening near the side wall is always fixed at 50μm. Figure 2a suggests that for our system, a ratio of 2:1 (i.e., 100μm: 50μm) has the lowest deviation or the best mixing efficiency, while 4:1 has the highest deviation or the worst mixing efficiency. Ideally, the diffusive mixing distance is proportionally related to $\sqrt{Dt}$, where D is the diffusion coefficient, and t is the residence time of the contacting fluid. Thus, a longer channel with many turns and returns can increase residence time (t), or mixing can be increased using a fluid with a higher molecular diffusion coefficient (D). As 'D' is constant for the FITC buffer, we can only vary the residence time. On the other hand, we wanted to use the area confined by the channel geometry. Moreover, we have yet to introduce two more variables into our design, so we preferred to choose a 1:1 ratio (50μm: 50μm), which would allow us to introduce more obstacles into the channel. We introduced arrays of bars as our next variables, and our simulation result (Figure 2b) suggested that 3 arrays of bars with a 1:1 ratio are the most optimum combination. From Figure 2b, we did not get better results for 4 arrays of bars even though we had more bars inside the channel. This can be attributed to the technical issues we faced in increasing the mesh density. Introducing 100µm circular obstacles after fixing the first two factors allows us to create more geometric constrictions inside the channel. Figure 2c suggests that 5 circular obstacles have the best result in terms of mixing the fluids with 32% efficiency (standard deviation 0.34, so mixing efficiency, E = (1-2σ) x 100 = 32%). This is expected because we have the maximum number of obstacles inside the channel. At this point, we hypothesize that the incoming streams fold up several times, diffusion distance stretches out, and we have better mixing with all these combinations (i.e., 1:1 ratio, 3 arrays of bars, and 5 circular obstacles). After optimizing all the variables, we ran our simulation at Re = 0.1 and Re = 10 to investigate the effect of lower and higher flow velocity, and the results are shown in Figure 2d. We observed the change in mixing efficiency down the length of the channel and plotted the results. At 5mm channel length, for Re ~0.1, the mixing efficiency (68%) is the highest according to Figure 2d. At this slow flow rate, the residence time is supposed to be the highest, and this result also suggests that the molecular diffusion length has the maximum width at this point, which is theoretically very hard to achieve. This could be verified by increasing the mesh density inside the channel and re-running the simulation.

**Methods**

To validate our simulation results, we utilized soft lithography and replica molding techniques to fabricate the mixer [30-32]. Soft lithography [33] is a simple, rapid, and inexpensive process of making 3D structures (master mold) to study a wide range of materials such as proteins and DNA. Replica molding is a widely used casting process of inversely replicating the master mold for making microfluidic channels. It readily duplicates the topological information of the master mold. This process enables rapid production as multiple copies can be formed from a single master mold. Figure 3 shows microscopic images of our master mold and fabricated device. We made the channels out of PDMS using replica molding [30-33]. At first, SU-8, a negative photoresist, was spin-coated on a pre-cleaned silicon wafer at a defined rotational speed. The spinning speed controls the thickness of the film on the wafer. After soft baking, a UV light exposes the desired area of the film through a photomask. Dissolving the unexposed area in a developer solution after the baking procedure produces the desired mixer patterns on the wafer. This is known as the master mold and

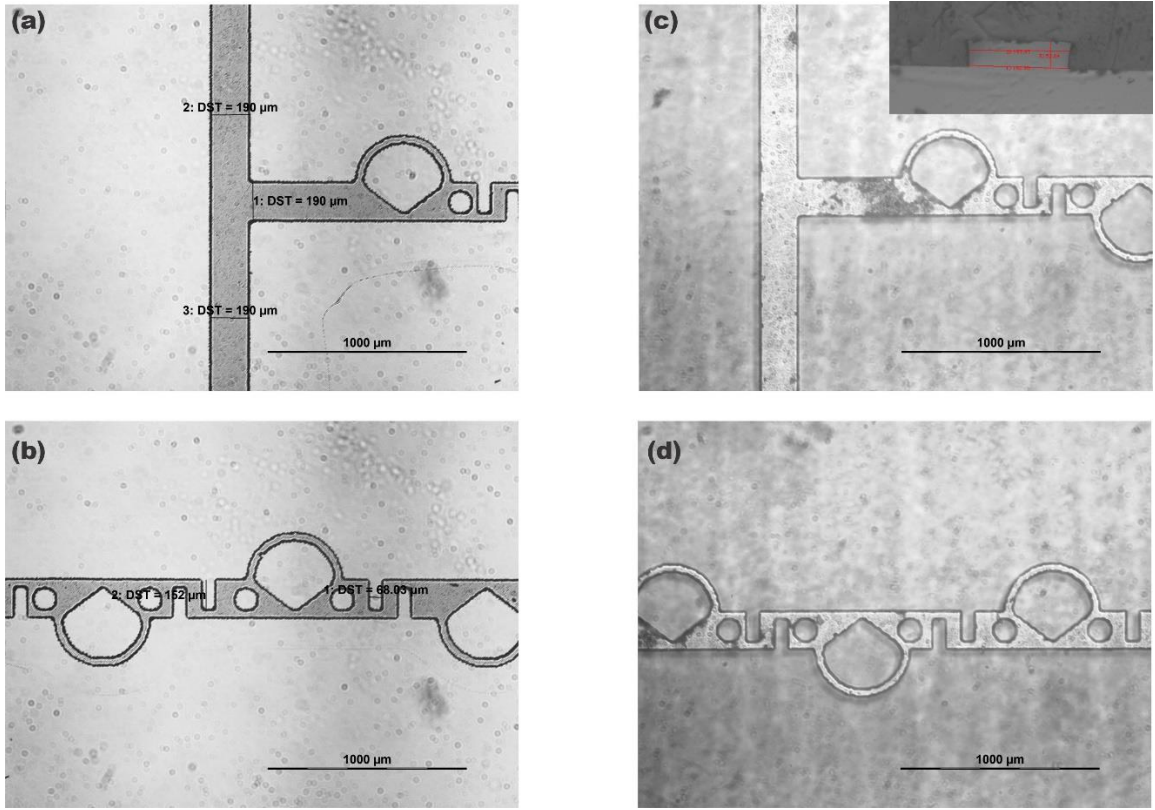

**Figure 3:** Top-down microscope images. (a) Top-down image of the master mold at the inlet. Channel width has been reduced by 5% (10µm). (b) Top-down image of the master mold that shows the features of the cone, rectangles, and circles inside the channel. (c) Fabricated device picture at the inlet. PDMS has picked dust near the first cone. In the inset, the cross-sectional view of the channel is shown where we have nearly 195µm width and 52µm height. (d) Top-down fabricated device picture of the mixer down at 3mm length.

Figure 3 (a and b) shows the top-down microscopic images of our master mold at two different locations of the channel. After hard baking, coating, and curing, the PDMS elastomer on this master mold produced an inverse replica of the topography, as illustrated by Figure 3 (c and d). The resulting inverse replica of the master constitutes the channel through which liquids may be injected to test mixing. Finally, microfluidic channels are produced by first exposing the replica to plasma treatment and sealing it to a glass substrate. A cross-section of our channel is also shown in the inset of Figure 3c.

**Results**

To evaluate the performance of this mixer, we used FITC buffer with DI water as our two fluid streams. The concentration of the buffer was set at 10 mol/m$^3$ in one of the inlets, while the velocity of two inlet streams was fixed at 0.013 m/s, which corresponds to Re ~1.

A syringe pump controls the volumetric flow rate of the two fluids, and connector tubing was used to pump liquids from the syringe to the fabricated device. A fluorescent microscope with a matched filter was used to observe the flow through the mixer visually, and then we captured fluorescent images every 1mm till the end of the mixer at 5mm. Figure 4a shows all the top fluorescence images. Then, we carried out a line scan through all these images to extract our experimental data. A line scan along the width of the channel allows us to measure the intensity of the light, which is related to the mixing of these two streams. A no-mixing region in the channel would be the brightest part in the line scan with maximum intensity, while the darkest one would be the fully mixed region. We matched the pixel count of these line scans with the width of the channel (1 pixel ~ 1.02 micron) and then analyzed only the channel region. All the intensities are normalized by the inlet maximum and minimum intensities [25]. A secondary normalization is necessary if the local maximum/minimum intensities at any location down the length of the channel are higher than the inlet maximum/minimum intensities. This is done by area normalization. All the measured results are plotted in Figure 4b.

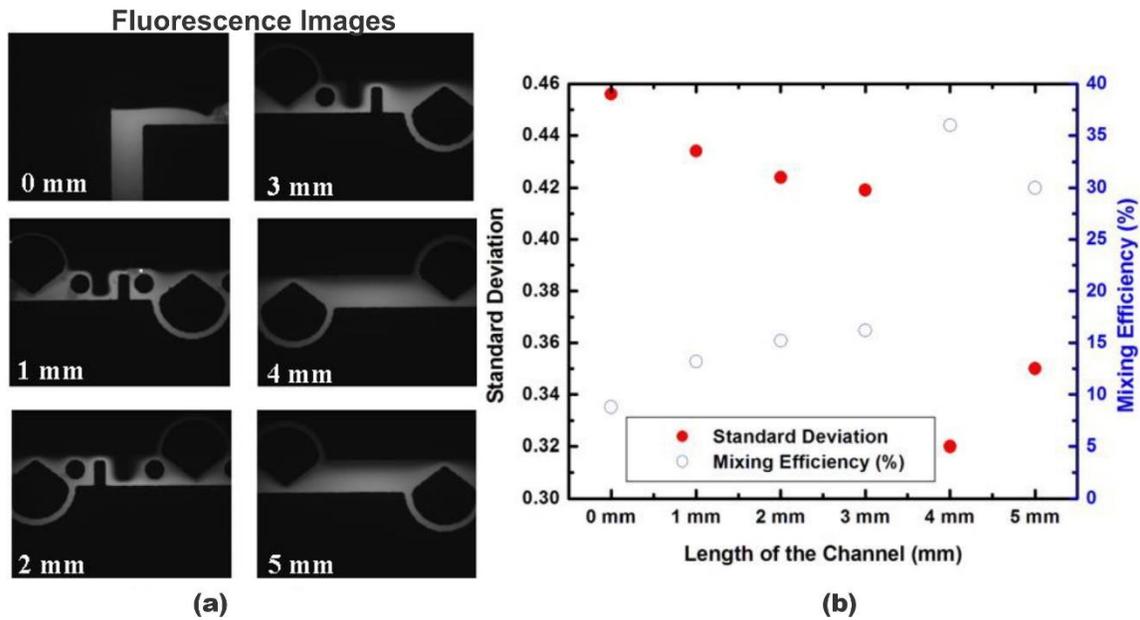

**Figure 4:** Experimental results of the proposed mixer. (a) Top-down fluorescence images of the device at 1mm separating the length of the channel. A syringe pump pumps FITC and Water, and images are captured at different locations. The intensities are directly related to concentration. Mixing width is measured after carrying out a line scan through all these images. (b) Standard deviation and mixing efficiency are plotted against the length of the channel. All the red-filled symbols are for the deviation, and open symbols are for the efficiency. Initially, the deviation is high, and down the length, it goes down, and the opposite thing happens for efficiency.

As we increase the number of obstacles inside the channel, our mixing efficiency shows an increasing trend until 4mm. From 4mm to 5mm channel length, the mixing efficiency has decreased by ~5.5%. We did not put any rectangular bars or circular obstacles in this position of the channel to improve mixing. We anticipated that the mixing efficiency at 4mm and 5mm would be the same. However, we found out that the mixing efficiency rather decreased. This could be attributed to three possible factors: a) time reversal occurred at this point, which decreased the residence time and, as a result, efficiency decreased; b) mass accumulation occurred due to misplacement of the outlet whole and, as a result of that backflow occurred inside the channel, c) wrong estimation of the intensity data due to mass accumulation. The measured result suggests that we have a 30% efficient mixer if we choose to use these variables inside a simple T-mixer.

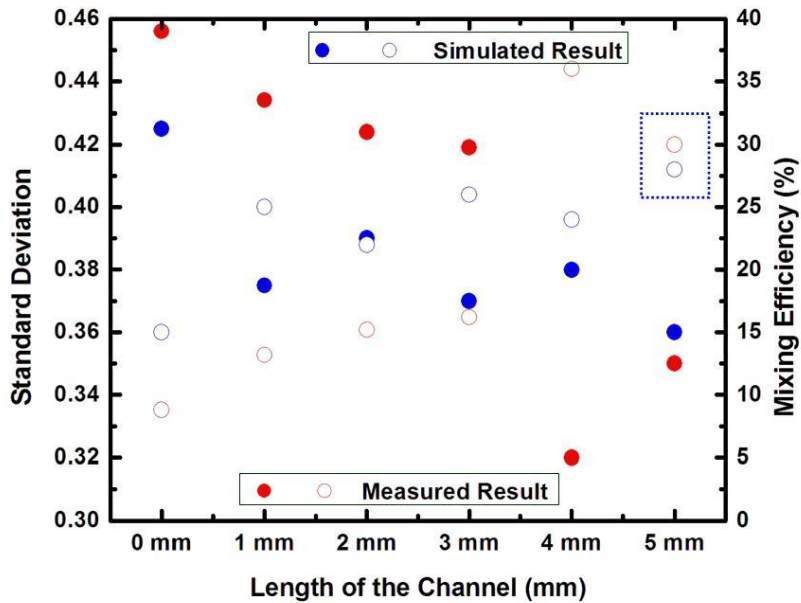

**Figure 5:** Comparison of the measured and simulated results. All the red colors are for the measured result, and the blue colors are for the simulated one. The closed symbols indicate standard deviation, and the open symbols indicate mixing efficiency. Both the simulation and mixing efficiency are in good agreement at the very end of the channel. Initially, there is a deviation between simulated and measured results. This is probably due to the assumption of ideal behavior during simulation.

Figure 5 compares the measured result with our simulation results as a function of increasing channel length. From 0mm to 3mm, our theoretical model predicted a less efficient mixer than what we found from the measurements. In a microfluidic channel, if 1D diffusion controls the mechanism of mixing, then more mixing can be predicted at the very end of the channel. We found this from our measured results from 0mm to 3mm. Our simulation results show some erratic nature from 1mm to 2mm of the channel length, but the measured results showed a nice decaying trend of standard deviation as we move from 0mm to 4mm. At 5mm, however, the mixing efficiency is actually lower than at 4mm. We have already pointed out a few factors for that. At 5mm, the mixing efficiency from simulation and measurement are in good agreement, which proves that our mixer is a well-behaved micromixer. The accuracy of our simulation results can be improved by changing a few parameters in COMSOL (e.g., mesh density).

**Conclusion**

Here, we presented a simple microfluidic mixer that achieved 30% mixing efficiency in a 5mm microfluidic channel. The design, fabrication, and characterization of this device are simple and easy to realize. We have not used any external source of energy to enhance molecular diffusion. It is, however, possible to improve the design of this mixer by using two or three separate units in series or using asymmetric obstacles inside the channel. Mixing, in general, is hard to achieve in a typical in-plane passive micromixer. Higher efficiency is generally obtained by increasing the length of the channel or increasing the number of geometric constrictions inside the channel so that flow splitting enhances mixing.

**Declaration**

This project was initiated as a class project by M. Rashed Khan in BME 590 (Biomedical Device Design) at NC State University, instructed by Dr. Glenn Walker. Device design, computation, experiments, and analyses were done by M. Rashed Khan for BME 590. Dr. Glenn Walker captured darkfield images and assisted M. Rashed Khan throughout as the course instructor of BME 590. Taliman Afroz is collaborating with M. Rashed Khan and Caleb Shaw for further extension and revision of the idea. Taliman Afroz and Caleb Shaw revised the original draft. This is an ongoing draft; modifications, revisions, and corrections are ongoing for the final submission to a peer-reviewed journal.